\documentclass[fleqn,usenatbib]{mnras}
\usepackage{newtxtext,newtxmath}

\usepackage[T1]{fontenc}

\DeclareRobustCommand{\VAN}[3]{#2}
\let\VANthebibliography\thebibliography
\def\thebibliography{\DeclareRobustCommand{\VAN}[3]{##3}\VANthebibliography}

\usepackage{graphicx}	
\usepackage{amsmath}	

\title[IGM Temperature Structure]
{The hydrodynamic response of small-scale structure to reionization drives large IGM temperature fluctuations that persist to $z = 4$}

\author[Cain et. al.]{
Christopher Cain,$^{1}$ \thanks{E-mail: clcain3@asu.edu}
Evan Scannapieco,$^{1}$
Matthew McQuinn,$^{2}$
Anson D'Aloisio,$^{3}$
and Hy Trac$^{4,5}$
 \\
$^{1}$School of Earth and Space exploration, Arizona State University, Tempe, AZ 85281, USA\\
$^{2}$Department of Astronomy, University of Washington, Seattle, WA 98195-1580, USA\\
$^{3}$Department of Physics and Astronomy, University of California, Riverside, CA 92521, USA\\
$^{4}$McWilliams Center for Cosmology and Astrophysics, Department of Physics, Carnegie Mellon University, Pittsburgh, PA 15213, USA\\
$^{5}$NSF AI Planning Institute for Physics of the Future, Carnegie Mellon University, Pittsburgh, PA 15213, USA\\
}

\date{Accepted XXX. Received YYY; in original form ZZZ}

\pubyear{2015}

\begin{document}
\label{firstpage}
\pagerange{\pageref{firstpage}--\pageref{lastpage}}
\maketitle

\begin{abstract}
The thermal history and structure of the intergalactic medium (IGM) at $z \geq 4$ is an important boundary condition for reionization, and a key input for studies using the Ly$\alpha$ forest to constrain the masses of alternative dark matter candidates.  Most such inferences rely on simulations that lack the spatial resolution to fully resolve the hydrodynamic response of IGM filaments and minihalos to HI reionization heating.  In this letter, we use high-resolution hydrodynamic+radiative transfer simulations to study how these affect the IGM thermal structure.  We find that the adiabatic heating and cooling driven by the expansion of initially cold gas filaments and minihalos sources significant small-scale temperature fluctuations.  These likely persist in much of the IGM until $z \leq 4$.  Capturing this effect requires resolving the clumping scale of cold, pre-ionized gas, demanding spatial resolutions of $\leq 2$ $h^{-1}$kpc.  Pre-heating of the IGM by X-Rays can slightly reduce the effect.  Our preliminary estimate of the effect on the Ly$\alpha$ forest finds that, at $\log(k /[{\rm km^{-1} s}]) = -1.0$, the Ly$\alpha$ forest flux power (at fixed mean flux) can increase $\approx 10\%$ going from $8$ and $2$ $h^{-1}$kpc resolution at $z = 4-5$ for gas ionized at $z < 7$.  These findings motivate more careful analyses of how the effects studied here affect the Ly$\alpha$ forest.  

\end{abstract}




\section{Introduction}
\label{sec:intro}

During the Epoch of Reionization (EoR), highly supersonic ionization fronts  (I-fronts) increased the temperature of the intergalactic medium (IGM)  by $1-3$ orders of magnitude \citep{Shapiro2004,Tittley2007,Venkatesan2011,DAloisio2019,Zeng2021}.  Measurements of the IGM temperature and its density dependence at $z \geq 4$ have begun to constrain the thermal history up to the Reionization epoch~\citep[][]{Becker2013,Walther2019,Boera2019,Gaikwad2020,Wilson2022}. These measurements constrain the reionization process itself~\citep[][]{Nasir2016,UptonSanderbeck16, Villasenor2022}, which likely ended at $z \approx 5.5$~\citep{Kulkarni2019,Keating2019,Nasir2020,Bosman2021}.  

The main observable used to probe the IGM thermal history is the HI Ly$\alpha$ forest~\citep[e.g.][]{Theuns2002}.  Ly$\alpha$ absorption is sensitive to the IGM gas temperature through both thermal broadening of the Ly$\alpha$ line and the temperature-dependent residual neutral fraction of ionized gas.  In the low-density IGM (densities with respect to the cosmic mean of $\Delta \lesssim 10$) and well after reionization, the temperature-density relation (TDR) is often well-approximated in reionization simulations~\citep{Kulkarni2015,Keating2018} by a power law of the form $T(\Delta) = T_0 \Delta^{\gamma-1}$,
where $T_0$ is the temperature at mean density and $\gamma - 1$ is the power law index~\citep{Hui1997, McQuinn2016}.  A number of studies have measured $T_0$ and $\gamma$ using the $2 < z < 5$ Ly$\alpha$ forest~\citep[e.g.][]{Becker2011,Boera2014,Hiss2018,Gaikwad2020}.  

Efforts to measure the thermal state of the IGM at $z > 4$ are important for several reasons.  First, the IGM thermal history at these redshifts is a boundary condition for reionization, and can help distinguish between reionization scenarios with different timings and durations~\citep[][]{Keating2019,Nasir2020}.  Second, much of the constraining power for constraining alternative dark matter cosmologies using the Ly$\alpha$ forest comes from $z\approx 4-5$ data ~\citep[e.g.][]{Viel2006,Viel2013,Baur2016,Irsic2017,Irsic2019}.  Such studies leverage the effects of the clumping properties of dark matter on small-scale structures in the Ly$\alpha$ forest~\citep[e.g.][]{Irsic2024}. These are sensitive to the IGM thermal history, so reliable inference requires reliable models for the thermal state of the gas.

A power-law TDR is motivated by heating/cooling processes that are important for low-density IGM gas.  These include heating from photoionization and cooling from the expansion of the universe and Compton scattering off the CMB~\citep{Hui1997,McQuinn2016}.  However, one process that has received relatively little attention is the effect of pressure-smoothing of the IGM by HI reionization on the dynamics of the IGM thermal structure\footnote{Although see~\citet{Puchwein2023} for a characterization of expanding filaments in the context of Ly$\alpha$ forest studies.  }.  After I-fronts sweep through a region, gas structures at mass scales $10^4 - 10^8 M_{\odot}$ undergo significant photoevaporation~\citep{Shapiro2004,Iliev2005b,Ciardi2006,DAloisio2020,Nasir2021,Chan2023}.  This causes expansion of dense gas initially trapped in filaments and minihalos, and compression in surrounding under-densities, which lead to adiabatic cooling and heating, respectively.  Capturing these processes requires spatially resolving the Jeans scale in cold ($\approx 10-1000$K), pre-ionized gas, which can be less than a ckpc~\citep{Gnedin2000}.  Such resolution is hard to achieve in most Ly$\alpha$ forest simulations, which require volumes $\geq (20 {\rm cMpc})^3$ to capture the relevant large-scale fluctuations~\citep{Doughty2023}.  Indeed, Ly$\alpha$ forest convergence studies have only begun reaching the scales necessary to resolve some of this effect\footnote{To our knowledge, the highest resolution achieved in any Ly$\alpha$ forest convergence study (at the mean density) is $5$ $h^{-1}$ckpc, in~\citet{Doughty2023}.  }.  Some studies have found evidence for convergence in Ly$\alpha$ forest simulations~\citep[e.g.][]{Bolton2009}, but often only in the limit that the IGM has been reionized for a long time, such that pressure smoothing takes place at redshifts much higher than are relevant for the Ly$\alpha$ forest.  

Several recent studies have begun to converge on the small-scale dynamics of IGM gas in the aftermath of cosmological I-fronts. 
A number of works have focused primarily on characterizing the destruction of minihalos and filamentary structures by reionization heating in the context of modeling the intergalactic ionizing opacity, and thus gave relatively little attention to the associated effects on temperature~\citep[e.g.][]{Park2016,DAloisio2020,Nasir2021,Chan2023}.  The question of how this process affects the thermal structure of the IGM (in the context of the Ly$\alpha$ forest) was first addressed by~\citet{Hirata2018} (see also~\citet{Montero-Camacho2024}).  They found that pressure-smoothing can induce significant and long-lived temperature fluctuations in the IGM, but only on the very small scales captured by their $\sim 300$ $h^{-1}$kpc boxes.  Here, we build on previous work in several crucial ways.  First, we address the numerical convergence of the gas physics and its effects on the Ly$\alpha$ forest, which has yet to be studied down to the $\sim$ kpc scales required to resolve pre-ionization clumping.  Second, we use simulation volumes large enough ($L_{\rm box} - 2$ $h^{-1}$Mpc) to capture a representative sample of cosmic structures while simultaneously capturing much of the post-ionization pressure-smoothing effect.  Finally, our simulations are performed with radiative transfer, which captures self-consistently the interplay between pressure smoothing, self-shielding, and the heating of IGM gas by I-fronts. 

In this letter, we study the effect of pressure smoothing on the thermal state of the IGM down to $z = 4$ in different reionization scenarios, and comment on possible implications for the high-redshift Ly$\alpha$ forest.  We use a suite of fully-coupled hydrodynamic+radiative transfer (RT) simulations of IGM gas dynamics at $z = 4-15$.  These have sufficient spatial resolution to capture much of the response of minihalos and filaments to the reionization process, and large enough volumes to at least capture some of the physical scales relevant for Ly$\alpha$ forest studies.   We describe our numerical methods in \S\ref{sec:numerical}, present our main results in \S\ref{sec:vis}, and conclude in \S\ref{sec:conc}.  Throughout, we assume the following cosmological parameters: $\Omega_m = 0.305$, $\Omega_{\Lambda} = 1 - \Omega_m$, $\Omega_b = 0.048$, $h = 0.68$, $n_s = 0.9667$ and $\sigma_8 = 0.82$, consistent with~\citet{Planck2018} results. Distances are in co-moving units unless otherwise specified. 

\vspace{-0.5cm}

\section{Numerical Methods}
\label{sec:numerical}

\begin{figure*}
    \centering
    \includegraphics[scale=0.21]{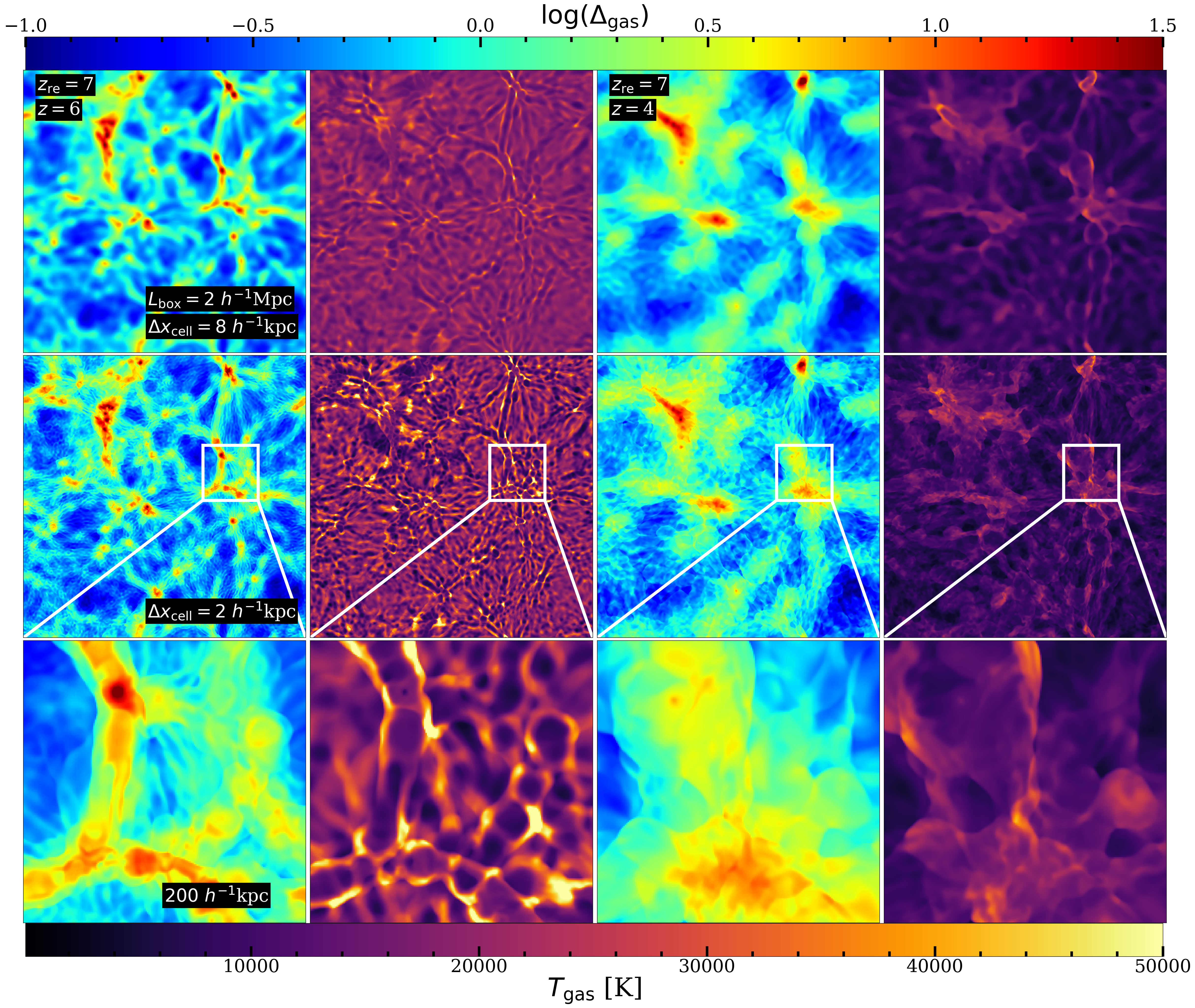}
    \caption{Visualization of the effects of pressure smoothing on the post-ionized IGM temperature.  The left half shows density and temperature slices at $z = 6$ from a high-res simulation re-ionized at $z_{\rm re} = 7$, and the right half shows the same for $z = 4$.  The top row shows a low-res simulation (with $8$ $h^{-1}$ kpc resolution), while the middle row shows a high-res ($2$ $h^{-1}$ kpc) simulation.  The bottom row shows a $200$ $h^{-1}$kpc zoom-in around an overdensity in the high-res box.  The dense gas inside the filaments is colder than its surroundings due to expansion cooling, while the surrounding gas has been heated by compression and (perhaps) weak shocks up to $\approx 50,000\,$K.  Even the low-density gas between filaments displays significant temperature fluctuations due to the expansion of local overdensities.  By $z = 4$ ($\Delta t \approx 780$ Myr), the expanding filaments have overlapped and cooled significantly, and the temperature structure more closely resembles that of the density field.  However, the effects of pressure smoothing (e.g. enhanced temperatures where expanding filaments overlap) remain conspicuous.  Comparison of the top and middle rows shows that these effects are much less pronounced in the low-res simulation, suggesting a substantial lack of convergence in the IGM thermal structure at this resolution -- a resolution similar to the highest-resolution simulations of the Ly$\alpha$ forest.  Indeed, noticeable differences persist to $z = 4$, suggesting that resolving these effects may be important for cosmological inference from the $z > 4$ Ly$\alpha$ forest.  }
    \label{fig:TD_vis}
\end{figure*}

We simulated IGM gas dynamics using a modified version of the RadHydro code presented in~\citet{Trac2004} and \citet{Trac2007}. RadHydro solves the RT equation with ray tracing, which is fully coupled to the gas dynamics on a uniform Eulerian grid.  Dark matter (DM) is treated in the Lagrangian frame and gravity is calculated using a particle-mesh scheme.  The simulations are initialized at $z = 300$ and run to $z = 4$.  The code uses a reduced speed of light approximation to speed up the RT calculation, and it solves for the chemical and thermal evolution of the gas using a sub-cycling backwards-difference solver.  All simulations are in $L_{\rm box} = 2$ $h^{-1}$Mpc boxes, and we have run simulations with $N = 1024^3$ DM particles, gas cells, and RT cells ($2$ $h^{-1}$kpc cells).   We also run $N = 256^3$ ($8$ $h^{-1}$kpc cell) simulations to study resolution convergence.  Our box size is chosen to get the largest possible statistical sample of cosmic structures whilst resolving most of the pressure smoothing effect.  Our lower resolution is close to the $10$ $h^{-1}$kpc resolution recommended by~\citet{Doughty2023}, which is on the high end of that achieved by most Ly$\alpha$ forest studies.  So, they likely capture as much thermal structure in low-density gas as previous Ly$\alpha$ forest simulations.  We refer to these as our ``high-res'' and ``low-res'' simulations, respectively.  

Our simulation setup is similar to that of~\citet{DAloisio2020}, designed to model the dynamics of IGM gas after it is heated by external ionizing sources.  We place $N_{\rm dom} = 16^3$ RT domains on a regular grid and send plane-parallel rays from all six faces into each domain.  Following~\citet{DAloisio2020}, the ionizing spectrum has a power law form, $J_{\nu} \propto \nu^{-\alpha}$ with $\alpha = 1.5$.  The radiation turns on everywhere at a specified redshift $z_{\rm re}$, with a flux density set to achieve a constant HI photo-ionization rate in optically thin gas, $\Gamma_{-12} \equiv \Gamma_{\rm HI}/10^{-12}$ s$^{-1}$.  In this study, we will consider simulations with $\Gamma_{-12} = 0.3$, which is close to what is measured from the Ly$\alpha$ forest at $4 < z < 6$~\citep[e.g.][]{Becker2013,DAloisio2018,Bosman2021}.   During the few Myr the radiation takes to cross the RT domains, we ``freeze'' hydrodynamics, gravity, and redshift evolution\footnote{Thermal evolution in cells less than $2\%$ neutral is also frozen during this time, which avoids un-wanted cooling effects.}.  We find that the gas temperature immediately following I-front passage ($T_{\rm reion}$) in our simulations is in good agreement with the model of~\citet{DAloisio2019} (their Eq. 3).  This guarantees that the hydrodynamic response of the gas to reionization is coherent everywhere in the box, with no un-physical gradients resulting from the RT domain structure.  The methods and simulations will be detailed in a forthcoming paper.

To calculate Ly$\alpha$ forest statistics, we draw $100$ random sightlines through the simulation volume that are long enough to cover the wavelength range of the Ly$\alpha$ forest ($1025-1215$ $\text{\AA}$, or $\sim 0.5$ $h^{-1}$Gpc).  At $z = 4$, this is a total path length of $\approx 40$ $h^{-1}$cGpc, sufficient to calculate converged Ly$\alpha$ forest statistics at scales captured by our boxes.  Note that since the boxes are periodic, a single sightline wraps around the box many times.  We include a condition on the angles of the random sightlines so that none of them point along a box axis, thus avoiding repeated structures. 

\section{Results}
\label{sec:vis}

Figure~\ref{fig:TD_vis} visualizes the effect of pressure smoothing on the IGM density and temperature.  The left and right halves show slices through the density and temperature at $z = 6$ and $4$ (respectively) for low-res (top row) and high-res (middle row) simulations re-ionized at $z_{\rm re} = 7$. The bottom panels show a $(200$ $h^{-1}$kpc$)^2$ zoom-in around an overdensity in the high-res simulation.  

At $z = 6$ ($\Delta t \approx 170$ Myr after ionization), the IGM is responding hydrodynamically to reionization.  The initially cold, dense filaments are being pressure-smoothed by photo-heating from reionization, driving compression of lower-density gas surrounding them.  The overlap of these expanding filaments generates an interference pattern in the density field that is conspicuous in the high-res simulations.  The temperature maps show the resulting complex thermal structure.  As filaments expand, their interiors are adiabatically cooled from $\approx 20,000-30,000$K to $\approx 5,000-10,000$K.  On the other hand, the gas on the edge of the densest filaments is heated by compression and possibly weak shocks up to $\approx 50,000$K~\citep{Chan2023}.\footnote{Through study of the entropy distribution of the gas, we find that most of the evolution immediately after reionization is consistent with adiabatic physics and not shocking, with only a small fraction of the gas showing significant entropy increases that evidence shocking.}   This is much larger than the initial post I-front temperature of $\sim 30000$K, and thus cannot be explained by radiative processes alone, which cool the gas at these temperatures.  This process results in filaments having a ``cored'' thermal structure~\citep[as noted by][]{Ocvirk2016}.  Finally, in the low-density gas surrounding filaments, milder, but still significant, $T$ fluctuations are driven by the expansion of smaller local overdensities ($\Delta \sim 1-5$).  As a result, even the low-density IGM displays a complicated thermal structure.  Note that our ``freezing'' procedure, described in \S\ref{sec:numerical}, guarantees that these effects are due to hydrodynamics and not gradients in the reionization redshift across the box (see also  Appendix~\ref{app:nyx} for tests with flash-ionized runs).  

The right panels show the same slice at $z = 4$ ($\Delta t \approx 780$ Myr).  By this time, the expansion of filaments is nearly complete and the gas has cooled significantly.  We see a stronger positive correlation between temperature and density than at $z = 6$.  This is because the processes that set the power law TDR have had enough time to start dominating the thermal structure of the gas.  However, pressure smoothing effects remain conspicuous in the temperature map.  Compressed gas at the boundaries of overlapping filaments is hotter than average, and the cored thermal structure of filaments is still visible.  This suggests that the imprint of pressure smoothing on the IGM thermal structure may persist well past the end of reionization.  Visual comparison of the top and middle rows shows that these effects, though still visible in the low-res simulation, are significantly under-resolved.  This is because these simulations lack enough resolution to capture the {\it pre-ionized} sizes of the filaments, and thus miss much of the pressure smoothing caused by reionization.  

\begin{figure*}
    \centering
    \includegraphics[scale=0.21]{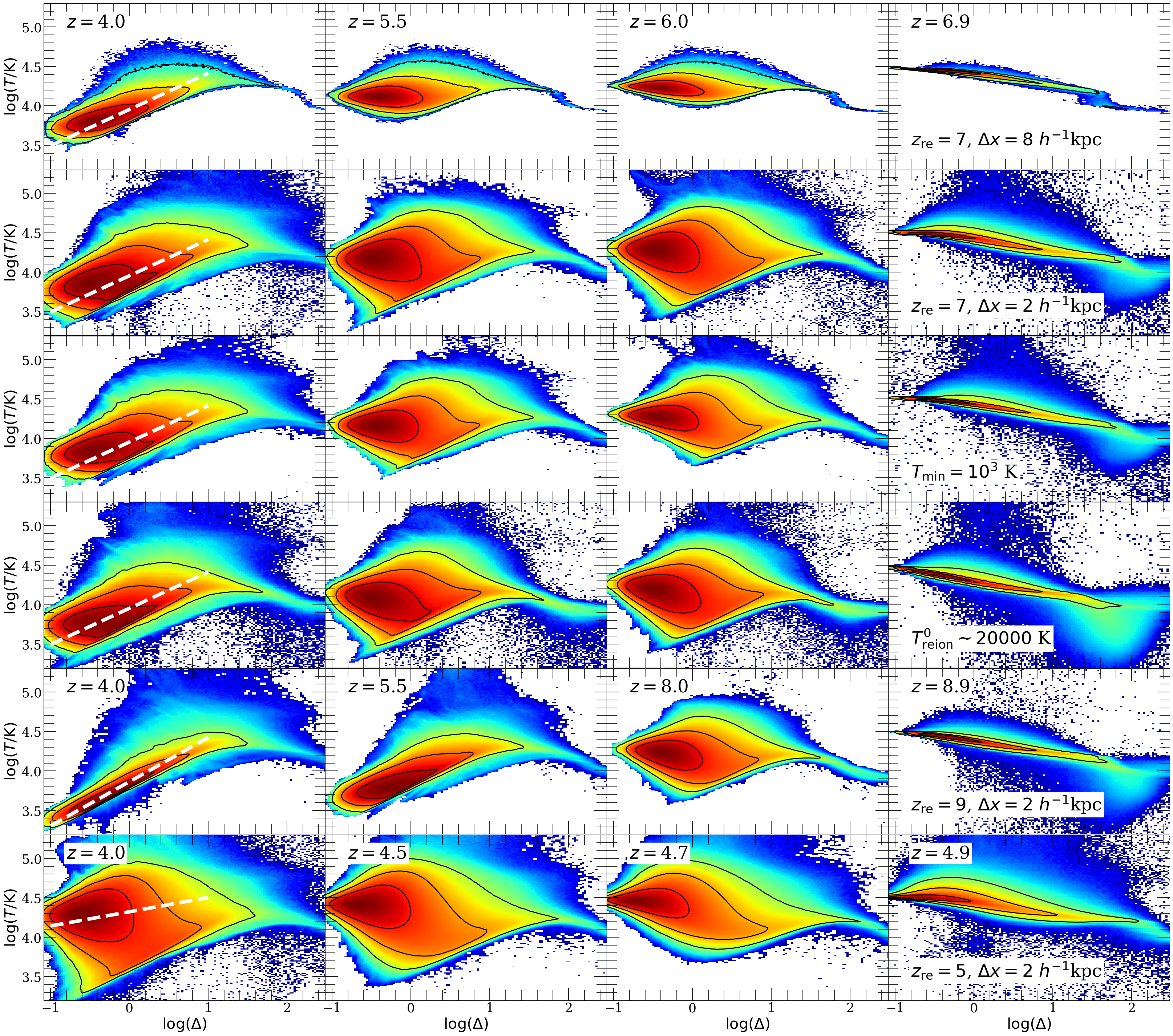}
    \caption{Effect of hydrodynamic response to reionization on the TDR.  Each panel shows the TDR in log color-scale, with the black lines denoting the $1$, $2$, and $3\sigma$ contours.  The top three rows show the TDR at several redshifts for  $z_{\rm re} = 7$.  The first two rows compare the low and high-res simulations shown in Figure~\ref{fig:TD_vis}, while the third row shows a high-res simulation in which we impose a minimum temperature floor of $T = 1000$K at $z < 15$, well before $z_{\rm re}$.  This temperature floor mimics the possible smoothing effects of X-ray preheating, which may lessen the subsequent effect from reionization.  The fourth row shows high-res results for $z_{\rm re} = 9$.  The bottom row shows the opposite extreme -- a patch that re-ionizes very late at $z_{\rm re} = 5$.   The white lines in the left column show the predicted TDR slope from the analytical model of~\citet{McQuinn2016}.  For $z_{\rm re} = 9$, $7$, and $5$, these slopes are $\gamma_{\rm pred} = 1.57$, $1.46$, and $1.18$, respectively.  Note that the low-res case is characteristic of the best resolutions obtained in previous Ly$\alpha$ forest studies.  }
    \label{fig:phase_diagram}
\end{figure*}

Figure~\ref{fig:phase_diagram} quantifies the evolution of the TDR.  The top three rows show the $T-\Delta$ phase diagram at $z = 4$, $5.5$, $6$, and $6.9$ (left to right) for simulations with $z_{\rm re} = 7$.  The 1st and 2nd rows show results for high and low-res simulations, respectively.  The 3rd row shows high-res results for a simulation with a temperature floor of $T_{\min} = 10^3$ K imposed at $z < 15$, well before $z_{\rm re}$.  This case is meant to roughly bracket possible effects of X-ray pre-heating on the pre-ionized structure of the IGM~\citep{Furlanetto2006,Fialkov2014c}.\footnote{X-ray pre-heating likely would happen more quickly at lower redshifts compared to our thermal floor implementation, leaving less time for pressure smoothing before reionization and, hence, the gas would be clumpier at a given pre-reionization temperature.}  Such pre-heating can increase the Jeans scale of pre-ionized gas, resulting in less initial clumping and less pressure-smoothing after I-fronts sweep through~\citep{DAloisio2020,Park2021}.  The 4th row shows the effect of $T_{\rm reion}$ on pressure smoothing.  We show a simulation with a factor of $10$ lower ionizing background of $\Gamma_{-12} = 0.03$ - in this run, the I-fronts moving much more slowly through the box initially, resulting in $T_{\rm reion} \sim 20000$K (at $\Delta = 1$).  The 5th row shows a high-res simulation with $z_{\rm re} = 9$ at $z = 4$, $5.5$, $8$, and $8.9$.  The bottom row shows a high-res sim re-ionized extremely late ($z_{\rm re} = 5$) at $z = 4.9$, $4.7$, $4.5$, and $4$.  The black lines denote $1$, $2$, and $3\sigma$ contours of the distribution.  We see that just after re-ionization (far right), the TDR is a tight, slightly inverted power law\footnote{The inversion results from high-density gas cooling for longer inside I-fronts, reaching a lower post I-front temperature.  }.  After $\Delta z = 1$, it evolves dramatically in the high-res runs.  The expansion/compression processes drive a $1-1.5$ dex scatter at $3\sigma$ in the temperature at mean density.  The effect is much less significant in the low-resolution run, which displays only $\approx 0.5$ dex scatter.  

Comparing the 1st and 2nd rows confirms our earlier observation: that {\it the effects of hydrodynamics on the TDR are highly un-converged in our low-res simulations}.  Since the initial sizes of filaments and minihalos are not captured in these runs, their hydrodynamic response to reionization is not fully captured either.  In fact, we have run a flash-ionized simulation with $1$ $h^{-1}$kpc resolution ($N = 2048^3$) with $z_{\rm re} = 7$ down to $z = 6$ and found that even our high-res runs are still mildly un-converged in this process (at least, in the limit of no preheating -- see Appendix~\ref{app:nyx}).  This lack of convergence suggests that even $2$ $h^{-1}$kpc resolution may not be sufficient to fully capture the effect under study.  Note that most of the scatter arises at densities $\Delta \lesssim 10$, which would not be star-forming gas - thus, the absence of star formation in our simulations should not affect this result.  The third row shows that the effect is somewhat reduced if IGM pre-heating is significant.  However, this case still displays much more thermal structure than the low-res run\footnote{We have also run simulations with $T_{\min} = 100$ and $10$ K.  We find only a small difference for the former and no appreciable difference for the latter.  }.   

The 4th row shows the effect of lowering $T_{\rm reion}^0$ (that is, $T_{\rm reion}$ at $\Delta = 1$) from $\approx 28000$K to $20000$K.  Based on Figure 2 of~\citet{DAloisio2019}, these correspond to I-front speeds of $\sim 2 \times 10^4$ and $2\times 10^3$ km/s for an $\alpha = 1.5$ spectrum, respectively, which roughly brackets the range expected during the latter half of reionization (see their Fig. 7, and also~\citet{Zeng2021}).  The TDR at $z = 6.9$ is inverted slightly more in this case.  The spread in the TDR at $z = 6$ and $5.5$ is slightly smaller than in the fiducial high-res case, but also slightly more than in the X-ray heated case.  At $z = 4$, the scatter is similar to the X-ray case and slightly less than the fiducial case.  Thus, we find that that differences in the I-front speed has an effect similar to (or less than) that of X-ray pre-heating.  

The 5th row demonstrates that, given enough time, the TDR does eventually ``relax'' to a power law with the expected slope.  The $z_{\rm re} = 9$ run displays the same behavior at $z = 8$ that the $z_{\rm re} = 7$ run does at $z = 6$, but by $z = 5.5$ the scatter is significantly lower, and at $z = 4$ ($\Delta t \approx 1$ Gyr from $z_{\rm re}$), a tight power law has been achieved with a scatter of only $\approx 0.3$ dex at $3 \sigma$.  However in the $z_{\rm re} = 7$ runs at $z = 4$ ($\Delta t \approx 800$ Myr), there is still significant scatter ($\approx 0.8$ ($0.6$) dex for the high (low)-resolution case and $\approx 0.7$ dex for the pre-heated model).  In the left-most row, the white dashed lines indicate the power law slope of the TDR at mean density expected from the analytical model of~\citet{McQuinn2016}.  The $z_{\rm re} = 9$ case closely follows this expectation, demonstrating that the heating/cooling processes responsible for it (e.g. photo-heating and Compton cooling) have largely erased the temperature fluctuations caused by pressure smoothing.  The bottom row ($z_{\rm re} = 5$) contrasts the $z_{\rm re} = 9$ scenario.  In this case, the hydro-driven fluctuations in $T$ peak at around $z = 4$.  It is unclear how long such a patch would take to reach a tight power law.  Patches of the IGM with $z_{\rm re} < 6$ may fill up to $20\%$ of the universe in realistic late-reionization scenarios~\citep{Kulkarni2019,Keating2019,Nasir2020}.  

These results may have complicating consequences for high-redshift Ly$\alpha$ forest studies that rely on concordance models of the IGM temperature.  Most Ly$\alpha$ forest simulations upon which such studies are based have resolutions of $\Delta x_{\rm cell} \geq 10$ $h^{-1}$kpc (at the mean density), which would likely cause them to miss much of the effect under study here.  In principle, this could affect efforts to interpret IGM temperature measurements parameterized by a power law, and/or inferences based on the Ly$\alpha$ forest~\citep[e.g.][]{Viel2013,Irsic2024} that do not fully account for these effects.  We emphasize that the effect of pressure smoothing on the IGM temperature structure is distinct from its effect on the density field itself, which has been well-understood and characterized in the context of the Ly$\alpha$ forest~\citep{Onorbe2016,Puchwein2023}.  It is also distinct from fluctuations on large-scales caused by the well-studied patchy reionization effect~\citep{Trac2008, 2015ApJ...813L..38D, Wu2019}.  Indeed, the degree of scatter in the TDR caused by the effect studied here is comparable to that found for patchy reionization by~\citet[][their Fig. 5]{Puchwein2023}.  However, it is unclear how these effects would interact since the pressure-smoothing effect has substantial $z_{\rm re}$ dependence.  

Since our simulations are large enough to capture Ly$\alpha$ forest statistics at some of the scales relevant for such studies, we have made a preliminary effort to quantify the importance of these effects on the Ly$\alpha$ forest power spectrum at fixed mean flux.\footnote{We re-scale our Ly$\alpha$ opacities such that the mean transmission matches measurements from~\citet{Becker2013} at $z = 4$ and \citet{Bosman2021} at $z = 5$.}  We find that in the extreme case with $z_{\rm re} = 5$, the $z = 4$ Ly$\alpha$ flux power spectrum at $\log(k /[{\rm km^{-1} s}]) = -1.0$ is $\approx 20\%$ higher in our high-res compared to our low-res simulations (with greater differences at larger $k$), suggesting significant lack of convergence.  For $(z_{\rm re}, z) = (7, 5)$, this discrepancy becomes $\approx 10\%$.  For $(z_{\rm re}, z) = (9,4)$ and $(7,4)$, we find $< 10\%$ disagreement between the low and high-res simulations.  These differences become significantly larger, though, when comparing to idealized scenarios in which the temperature-density relation is assumed to be a tight power law.  Note that the left column of Figure~\ref{fig:phase_diagram} shows that for all cases except $z_{\rm re} = 9$, the power-law parameterization represents the TDR poorly at $z = 4$.  Ly$\alpha$ forest studies aimed at dark matter constraints typically use $-2.5 \lesssim \log(k /[{\rm km^{-1} s}]) \lesssim -1.0$ in their analyses~\citep{Viel2013}, indicated that these differences could be important.  For Ly$\alpha$ forest analyses, these differences may be compensated by marginalization over thermal parameters during parameter inference.  However, the thermal model that is used does not capture most of the dispersion in the TDR found in our high-resolution simulations and so it is not obvious that this marginalization is sufficient to robustly constrain cosmological parameters such as the dark matter mass. 
We plan to follow up with a more detailed, Ly$\alpha$ forest-focused study.   

\vspace{-0.3cm}

\section{Conclusions}
\label{sec:conc}

In this letter, we studied the effect of pressure smoothing from reionization on the IGM temperature-density relation.  We found that the pressure smoothing of dense filaments and minihalos and compression of voids caused by this process results in a complex thermal structure that can persist to at least $z = 4$ in much of the IGM.  This structure differs substantially from the tight power law that is typically used to parameterize the thermal state of the low-density IGM in Ly$\alpha$ forest studies.  This effect is somewhat reduced, but still considerable, in simulations that assume significant pre-heating by X-ray sources prior to reionization.  We have demonstrated that simulations with $\geq 10$ $h^{-1}$kpc spatial resolution (at mean density), upon which most Ly$\alpha$ forest inferences rely, miss much of this process because they do not resolve the initial sizes of cold gas structures prior to reionization.  We have made a preliminary effort to quantify the effect on the Ly$\alpha$ forest flux power spectrum, estimating effects as large as several tens of percent -- with larger effects at higher $k$ and lower $z_{\rm reion}$.

Our preliminary results motivate several follow-up questions.  First, it is unclear how these effects would interact with the well-studied patchy reionization effect~\citep[e.g.][]{Trac2008,2015ApJ...813L..38D, Wu2019, Puchwein2023}.  We have found that the effects of pressure smoothing on temperature are sensitive to the time at which the gas was re-ionized.  It follows that this effect must be coupled to the large-scale patchiness of reionization.  This patchiness might also affect the Ly$\alpha$ forest at much lower wavenumbers than studied here, though we cannot address this question here with our small, single-reionization-redshift boxes.  Furthermore, our results might be affected by the inclusion of large-scale peculiar velocities, which are not well-captured in our small volumes and have been shown to be important for the Ly$\alpha$ forest~\citep[e.g.][]{Molaro2022}.  It is also unclear whether these effects could bias cosmological inferences from the Ly$\alpha$ forest.  This could be addressed in future work by combining small-box simulations at different $z_{\rm re}$ and box-scale densities (like the ones used in this work), and/or by achieving $\sim 2$ $h^{-1}$kpc in $10-20$ $h^{-1}$Mpc boxes.

\section*{Acknowledgements}

We thank George Becker and Vid Ir\v si\v c for helpful comments on the draft version of this manuscript.  CC acknowledges support from the Beus Center for Cosmic Foundations.  

\section*{Data Availability}

The data underlying this article will be shared upon reasonable request to the corresponding author.  



\bibliographystyle{mnras}
\bibliography{references} 

\appendix

\section{Comparison to Nyx \& Resolution Convergence}
\label{app:nyx}

\begin{figure}
    \centering
    \includegraphics[scale=0.19]{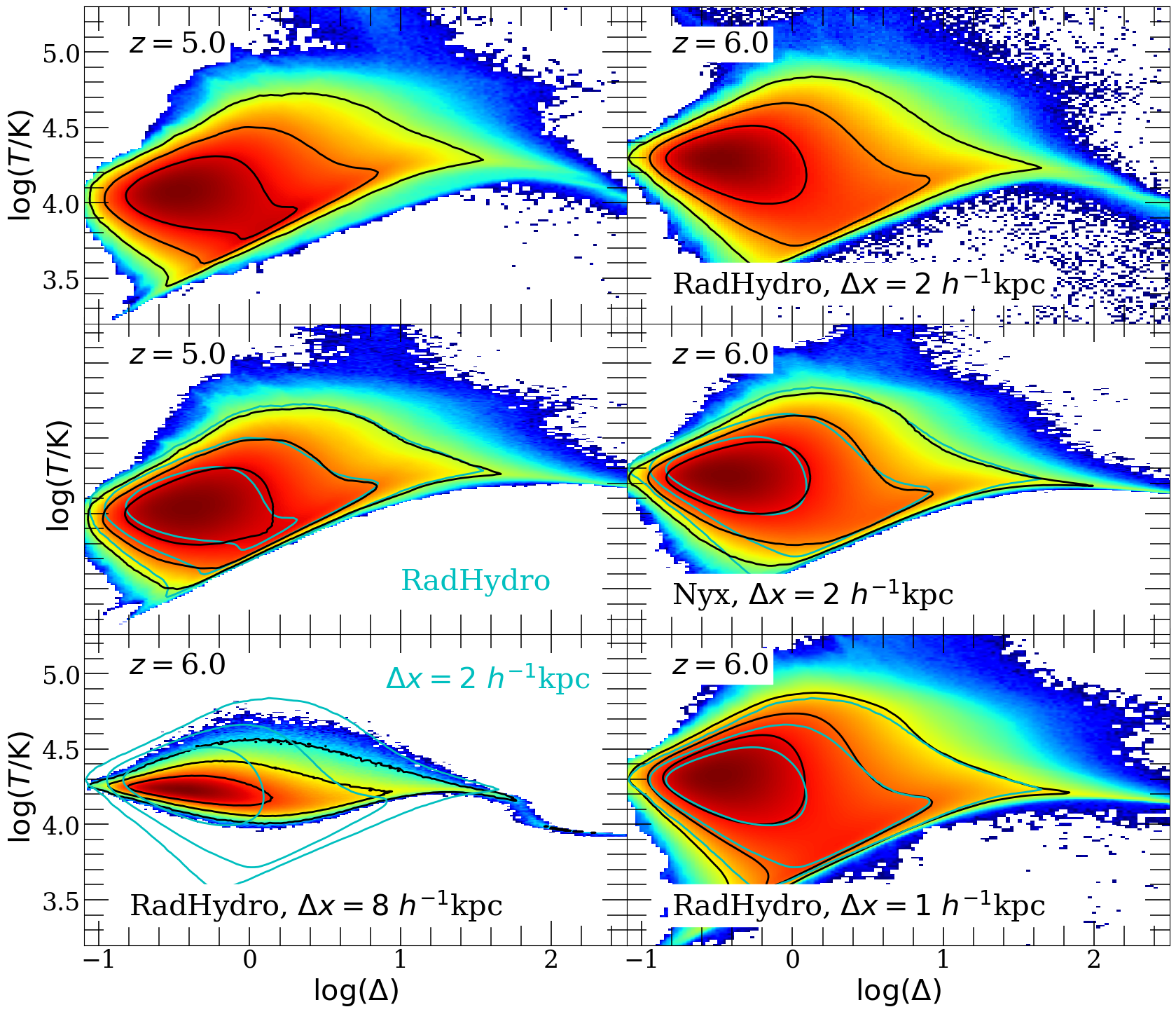}
    \caption{Comparison of the TDR in RadHydro and Nyx (top and middle row).  We flash-ionize a Nyx box at $z = 7$ and with $T_{\rm reion} = 32,000$K to codes at $z = 5$ and $6$ with $2$ $h^{-1}$kpc resolution.  In the Nyx panels, the cyan lines denote the corresponding RadHydro contours.  The agreement is excellent overall at both redshifts, the only notable difference being that RadHydro has a slightly wider distribution of temperatures near the mean density.  This is explainable by differences in the $T_{\rm reion}$ physics, as explained in the text.  The bottom row shows our low-res simulation (left) and our flash-ionized box with $1$ $h^{-1}$kpc resolution (right) at $z = 6$, with the cyan contours denoting our $2$ $h^{-1}$kpc run.  As seen in the main text, the former is dramatically un-converged.  The latter displays a slightly wider spread of temperatures than the $2$ $h^{-1}$kpc case (cyan contours), particularly at densities slightly close to the mean.  This suggests that even our high-res runs may not be fully converged (at least, in the limit of no pre-heating).  }
    \label{fig:nyx_phase_diagram}
\end{figure}

In this appendix, we compare our simulations with similar runs carried out with the Nyx cosmological hydrodynamic code~\citep{Almgren2013} and assess resolution convergence.  We have run Nyx simulations using the same initial conditions in our RadHydro runs with $2$ $h^{-1}$kpc resolution, with flash re-ionization at $z = 7$.  In Nyx, we can set the heat injection from reionization to produce a constant $T_{\rm reion}$.  We find that $T_{\rm reion} = 32,000$K reproduces the initial temperature of low-density gas in RadHydro reasonably well for $z_{\rm re} = 7$.  Note that this setup neglects the slightly lower $T_{\rm reion}$ values in filaments.  

The top two rows of Figure~\ref{fig:nyx_phase_diagram} show the TDR at $z = 5$ and $6$ for RadHydro and Nyx at our fiducial $2$ $h^{-1}$kpc resolution.  In the Nyx panels, the cyan lines show the corresponding RadHydro contours to aid the eye in comparison.  We find good agreement in the TDR between the two codes.  One difference is that the $T$ distribution close to the mean density is slightly wider in RadHydro, with more gas getting below $10^4$K.  This may be because the dense filaments start out colder in RadHydro than Nyx, since RadHydro accounts for the density dependence of $T_{\rm reion}$ (owing to its full RT treatment).  The densest structures ($\Delta \gtrsim 100$) are also able to self-shield against ionizing radiation for $10s$ to $100s$ of Myr after I-fronts pass through, and thus can remain cold long after the box ionizes.  

The bottom row shows the effect of spatial resolution.  The left and right panels show the TDR for our low-res simulation and that of a flash-ionized (no RT) run with $1$ $h^{-1}$kpc resolution.  The cyan curves denote the $2$ $h^{-1}$kpc-resolution contours.  As seen in the main text, our low-res runs are very un-converged in the TDR.  The $1$ $h^{-1}$kpc simulation displays a slightly wider spread of temperatures, particularly close to the mean density.  This suggests that even our high-res runs may not be fully converged in the behavior of the TDR on small scales.  This is un-surprising, since the characteristic sizes of pre-ionized structures can be $1$ $h^{-1}$ kpc or smaller if the gas is sufficiently cold.  However, we note that convergence criteria are likely to be most strict in these runs without X-Ray pre-heating as the gas near the mean density has cooled to $\sim 1$K by these redshifts (resulting in it being maximally clumpy).  Including the (highly uncertain) effects of pre-heating may eliminate the smallest gas structures and ease convergence criteria.  Thus, our convergence analysis reflects an upper limit on the resolution requirements of the TDR.

\bsp	
\label{lastpage}
\end{document}